\documentclass[5p,twocolumn,10pt,sort&compress]{elsarticle}

\journal{Astroparticle Physics}

\usepackage[latin1]{inputenc}         
\usepackage[T1]{fontenc}              
\usepackage[english]{babel}           

\usepackage{calc}
\usepackage[intlimits,namelimits]{amsmath}
\usepackage{amstext,amssymb,amsthm,mathrsfs}
\usepackage{fixmath}

\usepackage{latexsym}                 
\usepackage{textcomp}                 

\usepackage{float}                    
\usepackage{array}
\usepackage{color}
\usepackage{subfig}
\usepackage{graphicx}

\usepackage{setspace}                 
\usepackage{ellipsis}                 
\usepackage{url}                      
\usepackage{booktabs}                 
\usepackage{lastpage}                 
\usepackage{sectsty}
\usepackage{rotating}
\usepackage{verbatim}

\usepackage{units}

\usepackage{hyperref}

\newcommand{\td}{\,\mathrm{d}}	

\newcommand{\eqn}[1]{\label{eq:#1}}
\newcommand{\refeq}[1]{(\ref{eq:#1})}
\newcommand{\eq}{equation~\refeq}

\newcommand{\Eq}{Equation~\refeq}

\newcommand{\beq}{\begin{equation}}
\newcommand{\eeq}{\end{equation}}
\newcommand{\beqs}{\begin{eqnarray}}
\newcommand{\eeqs}{\end{eqnarray}}

\newcommand{\Leff}{\mathcal{L}_\text{eff}}
\newcommand{\Xe}{\text{Xe}}
\newcommand{\dedr}{\frac{\td\epsilon}{\td\rho}}
\newcommand{\Enr}{E_\text{nr}}

\begin{document}

\begin{frontmatter}

\title{\textbf{Interplay between scintillation and ionization \\ in liquid xenon Dark Matter searches}}
\author[a,b]{Fedor Bezrukov\fnref{fn1}}
\ead{Fedor.Bezrukov@physik.uni-muenchen.de}
\author[a,c]{Felix Kahlhoefer}
\ead{Felix.Kahlhoefer@mpi-hd.mpg.de}
\author[a]{Manfred Lindner}
\ead{Lindner@mpi-hd.mpg.de}

\address[a]{Max-Planck-Institut f\"ur Kernphysik, Postfach 103980, D-69029 Heidelberg, Germany}
\address[b]{Theoretische Teilchenphysik, Ludwig-Maximilians-Universit\"{a}t M\"{u}nchen, Theresienstr. 37, 80333 M\"{u}nchen, Germany}
\address[c]{Rudolf Peierls Centre for Theoretical Physics, University of Oxford, Oxford OX1 3NP, United Kingdom}

\fntext[fn1]{ASC Preprint Number: LMU-ASC 82/10}

\begin{abstract}
We provide a new way of constraining the relative scintillation efficiency $\Leff$ for liquid xenon. Using a simple estimate for the electronic and nuclear stopping powers together with an analysis of recombination processes we predict \emph{both} the ionization and the scintillation yields. Using presently available data for the ionization yield, we can use the correlation between these two quantities to constrain $\Leff$ from below. Moreover, we argue that more reliable data on the ionization yield would allow to verify our assumptions on the atomic cross sections and to predict the value of $\Leff$. We conclude that the relative scintillation efficiency should not decrease at low nuclear recoil energies, which has important consequences for the robustness of exclusion limits for low WIMP masses in liquid xenon Dark Matter searches.
\end{abstract}

\end{frontmatter}


\section{Introduction}

The recent results of XENON100~\cite{Aprile:2010um, Aprile:2011hi} correspond to a major increase of sensitivity for the Dark Matter (DM) searches and further improvements are expected soon when more data will be released. However, the analysis of the detector properties turns out to be subtle for low recoil energies, corresponding to low mass DM particles (below $\unit[10]{GeV}$).  The interest in this parameter region is additionally heated by claims of DM observation \cite{Bernabei:2008yi,Aalseth:2010vx}.  The problem is the proper reconstruction of the nuclear recoil energy from the primary scintillation signal (S1) in liquid xenon in the limit of low recoil energies.  Direct experimental calibration is rather difficult for low nuclear recoil energy and is prone to large systematic uncertainties, which led to mutually contradicting experimental measurements below $\unit[10]{keV}$ 
\cite{Arneodo:2000vc,Akimov:2001pb,Aprile:2005mt,Chepel:2006yv,Aprile:2006kx,Sorensen:2008ec,Aprile:2008rc,Manzur:2009hp,Lebedenko2009,Plante:2011hw,Horn:2011} (see figure~\ref{fig:severalleff}).

A theoretical treatment of the problem requires the determination of the scintillation yield for a slow moving xenon atom.  The task can be roughly divided into three parts: the problem of ionization and excitation probabilities in the individual collisions of xenon atoms, the problem of simulating the propagation through the media, and finally the problem of possible recombination of the produced free electrons and ions.  The outcome of such a theoretical treatment would be predictions for \emph{both} the scintillation and the ionization signals produced in the detector.

At sufficiently high energies the total electronic excitations in the atomic collisions can be reasonably well described by Lindhard's theory \cite{Lindhard}.  It approximates the process by point-like interactions between the incoming nucleus and electrons in the electron cloud of the target, and is applicable for the case of nuclear recoil velocity $v_\text{nr}\approx v_0 = e^2/\hbar$ (or $\Enr \approx \unit[1]{MeV}$). However, for smaller energies, the individual collisions are much harder to calculate and require a nonperturbative analysis of the electronic movement \cite{ISI:000187854300001}. For this reason, it is very difficult to describe inelastic collisions at energies much below $\unit[100]{keV}$.

In this article we do not attempt to make an ab-initio theoretical calculation of all the above processes, which is a very difficult task. Instead we make use of theoretical connections between the scintillation and ionization yields and the fact that the ionization yield is measured more reliably at low nuclear recoil energies than the scintillation yield \cite{Manzur:2009hp} (see figure~\ref{fig:severalqy}). 
For this purpose we will consider various possible modifications of the electronic stopping powers of liquid xenon using a simple parameterization
in order to compare the resulting scintillation and ionization yields with experimental data. From the data for the ionization yield we observe that we cannot choose stopping powers that lead to small electron excitations at small recoil energies without spoiling the fit of the ionization yield in figure~\ref{fig:severalqy}.  This observation leads to the prediction that the scintillation yield also cannot decrease much at small nuclear recoil energies. 

The paper is organized as follows: 
in section~\ref{sec:production} we review the process of generation of scintillation light and ionization in liquid xenon. Section~\ref{sec:stopping_power} introduces the notion of the nuclear and electronic stopping powers, which correspond to the analysis of individual xenon scattering events. Moreover, we calculate the total energy in the electronic excitations, and then analyze the recombination process in section~\ref{sec:recombination}. In section~\ref{sec:scintillation} all the results are collected and translated into the ionization and scintillation yields and compared with the experimental data. Section~\ref{sec:conclusions} contains our conclusions and further prospects.

\section{Production of scintillation light in liquid xenon}
\label{sec:production}

A large variety of effects must be taken into account to describe all physical processes that lead from the initial recoil to the production of scintillation light in liquid xenon. Specifically, we expect the following steps \cite{ISI:A1992JX92900021}:
\begin{itemize}
\item In a WIMP-like scattering event, an energy of 1--$\unit[100]{keV}$ is transferred to the nucleus.\footnote{Note that in the context of liquid noble gas detectors nuclear recoil energies are often quoted in keVnr. This unit is chosen to emphasize that any energy reconstructed from an S1 signal depends on the effective scintillation yield (and is therefore not necessarily physical). Since we are concerned with actual physical processes and not the detector signals in this paper, we will give nuclear recoil energies in keV.} As the corresponding recoil velocity is well below the Fermi velocity of the most loosely bound electrons, we expect the atom to remain neutral in the scattering process.
\item The recoiling atom will scatter off neighboring nuclei. While most scattering events are expected to be elastic, there will occasionally be inelastic collisions leading to excitation or ionization of either (or both) of the atoms.
\item After each scattering process both atoms will continue their propagation with a fraction of the initial recoil energy. Consequently, both can again scatter elastically or inelastically off other atoms.
\item During the process of thermalization the recoiling xenon atoms will leave behind a large number of ionized or excited xenon atoms~--- distributed along many branches of the initial track.
\item The free electrons will now either recombine with surrounding ions to form excited xenon atoms or escape from recombination. The fraction of escaping electrons will depend on the strength of the applied electric drift field, but some electrons will escape even in the absence of a field.
\item Excited xenon atoms are free initially, but will soon be self-trapped and form excimers. These excimers emit vuv scintillation light on the transition to the ground state. In a simplified picture, the process is
\beqs 
\Xe^* + \Xe & \rightarrow & \Xe_2^* \\
\Xe_2^* & \rightarrow & 2\Xe + h\nu
\eeqs
\item In some cases, especially at high excitation density, two exited xenon atoms will combine to produce only one scintillation photon. This process, known as biexcitonic quenching, will effectively reduce the scintillation yield.
\end{itemize}

At first sight, the large number of steps seems to make it very hard to disentangle possible ambiguities. A decrease of scintillation efficiency at low recoil energies could, for example, be equally attributed to a decreasing cross section for inelastic scattering (for example due to threshold effects), a different track structure, an increasing fraction of escaping electrons or a stronger quenching mechanism. This ambiguity can be lifted at least partially by considering not only the effective scintillation yield of nuclear recoils, but also the ionization yield (see also~\cite{Aprile2010RvMP}). This quantity is much better measured but has been~--- to the best of our knowledge~--- ignored in all previous attempts to give a theoretical model for the scattering process in liquid xenon.

The sum of ionization and scintillation, which we will refer to as the total electronic excitation, should correspond to the total energy lost in inelastic collisions. Consequently, it should only depend on the scattering cross sections and not on the processes occurring later, such as recombination, which will only lead to a redistribution between ionization and scintillation. Thus, if both signals showed a similar energy dependence, this would suggest a general suppression of inelastic scattering at low energies~--- which is what one might naively expect. However, one actually observes experimentally a strong increase of the ionization yield at low energies. This observation indicates that the suppression of the scintillation signal at low energies does not result from the actual inelastic scattering processes, but from the large number of escaping electrons.

The effective scintillation yield of nuclear recoils is usually described by the dimensionless quantity $\Leff$, called relative scintillation efficiency. It relates the S1 scintillation signal to the physical recoil energy of the nucleus $E_\mathrm{nr}$ as
\beq E_\mathrm{nr} = \frac{S1}{L_\mathrm{y} \cdot \Leff} \cdot \frac{S_\mathrm{e}}{S_\mathrm{n}} \;. \eeq 
$L_\mathrm{y}$ is the light yield for $\unit[122]{keV}$ electron recoils and $S_\mathrm{e,n}$ are the electric field quenching factors for electronic and nuclear recoils. Thus, $\Leff$ quantifies the suppression of scintillation for nuclear recoils compared to $\unit[122]{keV}$ electron recoils at zero electric field.

\section{Stopping powers of liquid xenon}
\label{sec:stopping_power}

In this section we will describe the interactions of neutral xenon atoms and discuss possible scattering processes at energies of a few keV. The quantities we are interested in are the rate at which energy is transferred to recoiling nuclei by elastic collisions and the rate at which electrons are excited by inelastic collisions. These quantities are often called nuclear stopping power and electronic stopping power.

\subsection{Electronic stopping power}
\label{sec:electronic_stopping}

The electronic stopping power is defined as the average energy which an atom loses to electronic excitations per distance travelled through the detector, $\left(\td E / \td x\right)_\mathrm{e}$. In a ``semiclassical'' approach, an electron is excited when it collides with a nucleus. The stopping power should therefore be proportional to the electron mass density $n_0$, their velocity $v_\mathrm{F}$, the momentum transfer cross section $\sigma_\mathrm{tr}(v_\mathrm{F})$, and the velocity of the incoming particle $v$. In fact, electronic stopping is often described by \cite{ISI:A1995QT40100056, ISI:A1994MW35200061, ISI:A1986AYM1800013, ISI:A1977DP88300015, ISI:000184186600054}
\beq \left(\frac{\td E}{\td x}\right)_\mathrm{e} = n_0 v v_\mathrm{F} \sigma_\mathrm{tr}(v_\mathrm{F})\;. \eqn{elstop}\eeq

We should take a moment to discuss the validity of such a semiclassical approach. \Eq{elstop} is in fact based on several assumptions:
\begin{itemize}
\item Instead of describing the atomic system by a many-particle wavefunction, we claim that only the electron density is relevant to the problem, \emph{and} we ignore the modification of the electron density during the atomic collision.
\item Collisions between electrons and nuclei are treated like classical point like interactions.
\item Electrons are assumed to be free. Consequently, no minimal energy transfer is required for an excitation.
\end{itemize}

For a uniform electron gas the electronic stopping power is proportional to the velocity of the incoming particle, a result that has been derived in~\cite{Lindhard}. The authors obtain
\beq \left(\frac{\td E}{\td x}\right)_\mathrm{e} = \sqrt{8} \pi e^2 a_0 \zeta_0 Z N \cdot \frac{v}{v_0}\;, \eqn{Lindhard_stopping} \eeq
where $a_0$ and $v_0$ are Bohr radius and Bohr velocity respectively and $\zeta_0$ is an empirical parameter, often\footnote{In an independent derivation, Firsov~\cite{Firsov} obtained the same formula with $\zeta_0 \approx 1.63$.} set to $\zeta_0 = Z^{1/6}$. $N = \unit[13.76]{nm^{-3}}$ is the number density of xenon atoms. This formula should also be valid for nuclear velocities $v < v_0$.

However, there is no reason that we can extrapolate the velocity proportional behavior all the way down to $v = 0$.
In fact, at very low velocities, departures from velocity-proportionality have been observed experimentally \cite{ISI:A1994MW35200061}. Moreover, there are several theoretical arguments in favor of a more rapid drop of $\left(\td E/\td x\right)_\mathrm{e}$ for $v<v_0$.  The two most important ones are threshold effects and Coulomb effects.

The argument for threshold effects essentially goes as follows: in an elastic collision between a nucleus and a free
electron, only a fraction $m_e / m_N$ of the nucleus energy can be transferred to the electron. For nuclei with energies in the keV range, the resulting electron energy is at most a few eV~--- so we can no longer ignore gap energies or the work function of xenon. There has been a long and intense discussion on whether such threshold effects are present or not (see for example \cite{Ficenec:1987vy}). However, many experiments and theoretical considerations report electronic excitations far below the naive threshold \cite{ISI:A1975W255400016, ISI:000187854300001}. 

An effect from Coulomb repulsion is expected, because at very low relative velocities colliding nuclei will not penetrate the electron clouds of each other strongly. Consequently, with decreasing energy the recoiling nucleus will probe only regions of lower electron density \cite{ISI:A1986A763000023}.

Both of these arguments are doubtful, because they continue to exploit the point-like interaction of the nucleus with the electron.  However, for low nuclear velocity the electron clouds rearrange during the collision (or, in a more semiclassical language, the electron makes several rotations in the combined electric field of the two colliding atoms during the collision).  This effect leads to a much more complicated non-perturbative mechanism of the energy transfer to the electron. For an analysis of such collisions in the case of simple atoms, we refer to \cite{ISI:000187854300001}.

Ideally, exact or approximate quantum mechanical calculations for the Xe--Xe scattering process must be performed.  Lacking such calculations, we will continue to use \eq{Lindhard_stopping} even for low energies. However, we will introduce a correction factor $F(v/v_0)$ to parameterize our ignorance of the cross section:
\beq \left(\frac{\td E}{\td x}\right)_\mathrm{e} \to F(v/v_0) \left(\frac{\td E}{\td x}\right)_\mathrm{e}\;. \eqn{suppfactor}\eeq
In the end, we will compare our results with the experimental data on ionization and scintillation for different choices of $F(v/v_0)$. 
Until then, we will use $F(v/v_0) = 1$ unless explicitly stated otherwise.

To conclude this section, we provide dimensionless quantities instead of $\left(\td E / \td x\right)_\mathrm{e}$, which are usually preferred in the literature. Therefore, we define the reduced energy $\epsilon$, reduced distance $\rho$ and a dimensionless electronic stopping power $s_e$ by
\beqs \epsilon & =  & \frac{a}{2 e^2 Z^2}E \\ 
\rho & =  & N \pi a^2 x \\ 
s_e & = & \dedr = \frac{\left(\td E / \td x\right)_\mathrm{e}}{2 \pi e^2 a Z^2 N} = \frac{a_0 \zeta_0}{a} \sqrt{\frac{8\epsilon}{e^2 a m_N}} \eqn{sered} \;, \eeqs
where $a = 0.626 a_0 Z^{-1/3}$ is the Thomas-Fermi screening length. For liquid xenon, the reduced energy $\epsilon$, which we will use throughout the rest of the paper, can be expressed as $\epsilon = \unit[1.05 \cdot 10^{-3} E_\mathrm{nr} /]{keV}$.

\subsection{Nuclear stopping power}
\label{sec:nuclear_stopping}

The second quantity needed to calculate the amount of energy lost to electronic excitations is the nuclear stopping power, corresponding to the probability for elastic scattering of two xenon atoms. To calculate the cross section, we approximate the electron wave functions by the electron density, and ignore modifications of the electron clouds during the collision. The energy transfer in such a collision depends in general on the energy $E_\mathrm{nr}$ of the projectile and the scattering angle $\theta$. However, it turns out that due to scaling properties all relevant functions depend only on the combined variable \cite{ISI:000224129400001}
\beq \eta = \epsilon \sin \frac{\theta}{2} \;, \eeq
where $\epsilon$ again denotes the reduced energy. 

The differential cross section for elastic scattering can then be written as
\beq \frac{\td\sigma}{\td\eta} = \pi a^2 \frac{f(\eta)}{\eta^2} \; ,\eeq
where $a$ is again the screening radius. The function $f(\eta)$ depends on the screening function that we adopt to describe the charge density. For a large number of screening functions, $f(\eta)$ can approximately be written as \cite{Winterbon}
\beq f(\eta) \approx \frac{\lambda \eta^{1-2m}}{\left(1+\left[2\lambda\eta^{2(1-m)}\right]^q\right)^{1/q}} \;. \eeq
From $f(\eta)$ we can obtain the dimensionless nuclear stopping power $s_n(\epsilon)$:
\beq s_n(\epsilon) = \frac{1}{\epsilon}\int_0^\epsilon \td \eta f(\eta) \;. \eeq

\begin{figure}[tb]
\centering
\subfloat[Screening function]{\includegraphics[width=0.8\columnwidth]{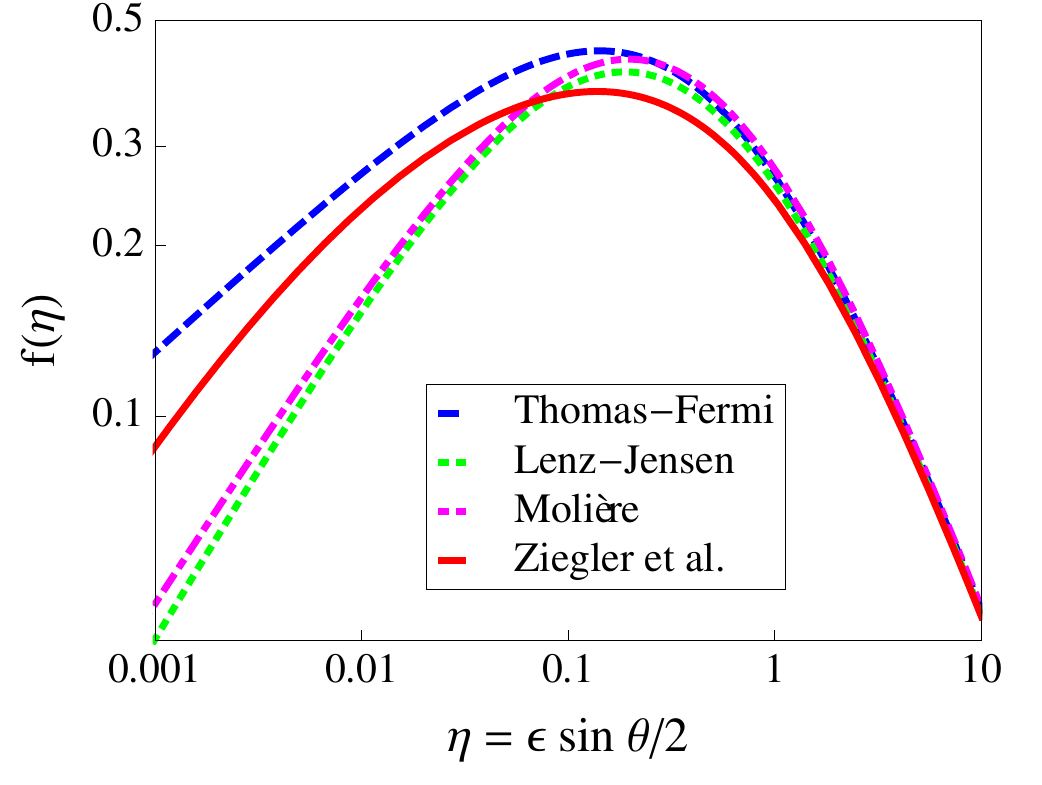}}
\qquad
\vspace{3mm}
\subfloat[Nuclear stopping power]{\includegraphics[width=0.8\columnwidth,clip,trim=7 0 60 0]{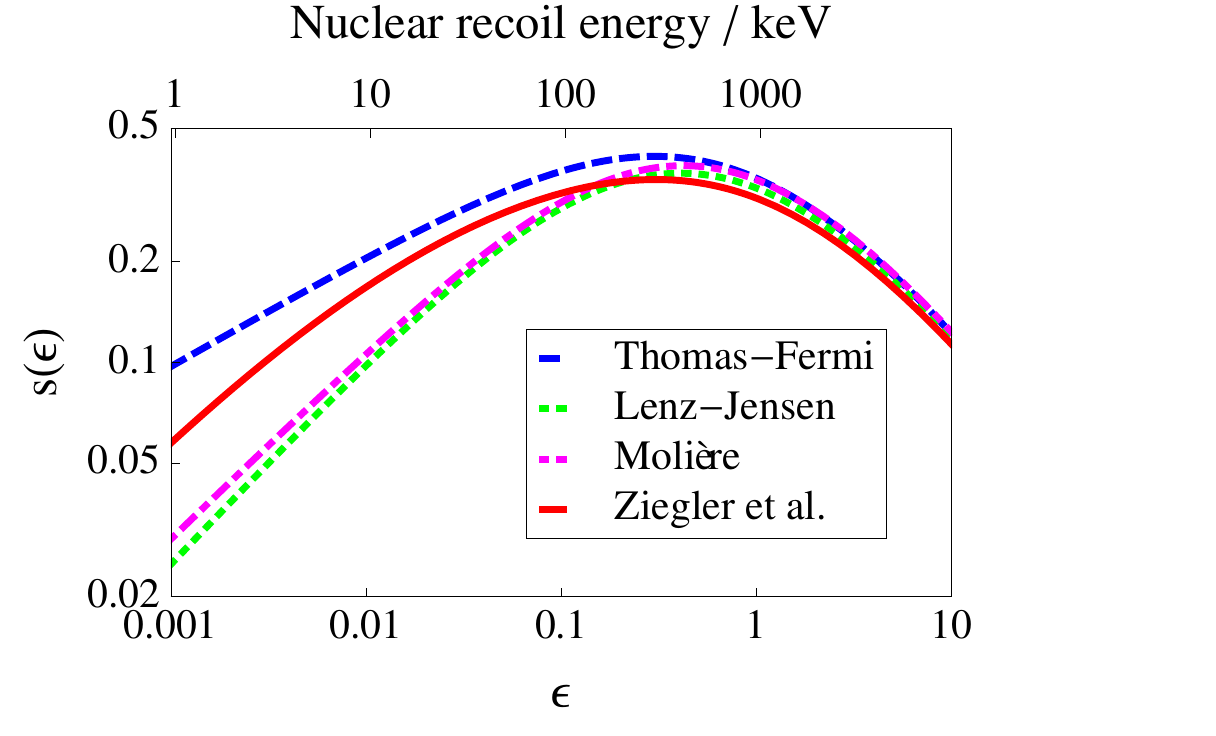}}
\caption{Comparison of different choices for the nuclear stopping power.}
\label{fig:comparision_stopping}
\end{figure}

Both quantities, $f(\eta)$ and $s_n(\epsilon)$ have been calculated by various authors with differing results. These differences, especially in the low energy region, are due to different approximations for the screening function. Lindhard et al.\ favor the Thomas-Fermi screening function corresponding to $m = 0.333$, $q=0.667$, $\lambda=1.309$. However, today it is generally agreed that the Thomas-Fermi screening function overestimates the potential at large distances and therefore gives too large stopping powers at low energies \cite{ISI:000224129400001}. One therefore often prefers the Moli\`{e}re or the Lenz-Jensen screening functions that show better agreement with experimental data. They correspond to the parameter choices $m = 0.216$, $q=0.570$, $\lambda=2.37$ and $m = 0.191$, $q=0.512$, $\lambda=2.92$, respectively. For a comparison of the different screening functions, see figure~\ref{fig:comparision_stopping}.

The most reliable way to calculate the screening function is to use Hartree-Fock methods instead of an analytical approximation. Ziegler et al.~\cite{INSPEC:2666373} find the best agreement with experimental data for the (so-called universal) screening function given by the following expression:\footnote{It is important to notice that Ziegler et al.\ use a slightly different definition for the reduced energy, because they assume a different screening length. For xenon, the conversion factor is $\epsilon_\mathrm{Z} = 1.068 \epsilon$.}
\beq s_n(\epsilon_\mathrm{Z}) = \frac{\ln(1+1.1383 \epsilon_\mathrm{Z})}{2\left[\epsilon_\mathrm{Z} + 0.01321 \epsilon_\mathrm{Z} ^ {0.21226} + 0.19593 \epsilon_\mathrm{Z} ^{0.5}\right]} \; . \eqn{nucstop} \eeq
From this, $f(\eta)$ can be calculated using $f(x) = \frac{\td}{\td x}\left[x s(x)\right]$.

In contrast to the universal screening function from Ziegler et al.\, the advantage of the simpler screening functions is that $f(\eta)$ can be approximated by a power law for very small values of $\eta$ (i.e.\ $\eta < 10^{-4}$):
\beq f(\eta) \simeq \lambda \eta^{1-2m} \; .\eqn{f_scaling} \eeq
As we will not need this property, we will use the universal screening function in the following, because it is the most accurate at low velocities and agrees best with experimental data. Note that the uncertainty related to the nuclear stopping power can be absorbed into the correction factor introduced in \eq{suppfactor}, so that our discussion remains general. In figure~\ref{fig:stopping}, we will show how our final results would be affected by making a different choice for the screening function.

\subsection{Total electronic excitation}
\label{sec:Lindhard}

In this section we will combine the nuclear stopping power $s_n(\epsilon)$ from \eq{nucstop} and the electronic stopping power $s_e(\epsilon)$ from \eq{elstop} in order to predict what amount of the initial recoil energy is transferred to electronic excitations. We will denote the total energy in electronic excitations by $\kappa(\epsilon)$ and also define the quotient
\beq \xi(\epsilon) = \frac{s_e(\epsilon)}{s_n(\epsilon)} \; . \eeq

First of all we should try to discuss the general trends which we expect for $\kappa(\epsilon)$ and $\xi(\epsilon)$. At energies above one MeV, inelastic collisions will dominate because the electronic stopping power grows proportional to $\sqrt{E}$, while the nuclear stopping power decreases after reaching its maximum around $\unit[100]{keV}$. Consequently, $\kappa(\epsilon) \approx \epsilon$ in this energy region. At energies of a few keV and below, the nuclear stopping power is much larger than the electronic stopping power. Consequently, we expect $\kappa(\epsilon) \ll \epsilon$ and $\xi(\epsilon) \ll 1$.

In the low energy region, a first estimate for $\kappa(\epsilon)$ is given by
\beq \kappa(\epsilon) = \frac{\epsilon \cdot s_e(\epsilon)}{s_n(\epsilon) + s_e(\epsilon)} \approx \epsilon \cdot \xi(\epsilon) \;. \eqn{kappa_estimate} \eeq
Making this estimate, we have included only electronic excitations produced by the primary recoiling nucleus. We expect, however, that at least some recoiling nuclei from secondary elastic collisions still have enough energy to inelastically excite other atoms. Thus, part of the energy lost in elastic collisions can still be transferred into electronic excitations. 

One way to take this effect into accout would be to perform numerical simulations. Here, we will instead rely on Lindhard et al.\ \cite{ISI:A19611456C00030}, who have derived an integral equation to determine $\kappa(\epsilon)$. Under the approximation that most electronic excitations occur at large impact parameter and have only small energy transfer, the authors show that for $\xi(\epsilon) \ll 1$, $\kappa(\epsilon) \propto \epsilon \cdot \xi(\epsilon)$. 
Thus, we only need to modify our simple estimate in \eq{kappa_estimate} by introducing a constant of proportionality $\alpha$ in order to include the effect of secondary nuclear recoils:
\beq \kappa(\epsilon) = \alpha \epsilon \xi(\epsilon).\eqn{kappa}\eeq 
This constant of proportionality can be calculated analytically, if $\xi(\epsilon)$ can be described by a power law. As we want to consider cases where $\xi(\epsilon)$ cannot be described by a power law, we cannot calculate $\alpha$. However, we will show that this presently unknown constant can be absorbed into another quantity which we can determine from experimental data. Since we expect the simple estimate in \eq{kappa_estimate} to underestimate $\kappa$, we require $\alpha > 1$, but still of order 1, for consistency.

It should be emphasized that \eq{kappa} does take into account secondary recoils produced along the entire track of the primary nucleus. Its simple structure is a result of the assumption $\epsilon \ll 1$, which implies that the nuclear stopping power is much greater than the electronic stopping power and both decrease rapidly as the recoiling nuclei loose their energy. More accurate ways to calculate $\kappa$ have been considered in the literature (see for example~\cite{Mei:2007jn}). However, the original result from Lindhard's theory has the virtue that $\kappa \propto F(v / v_0)$, which will simplify the subsequent analysis.

\begin {figure}
\begin{center}
  \includegraphics[width=0.8\columnwidth]{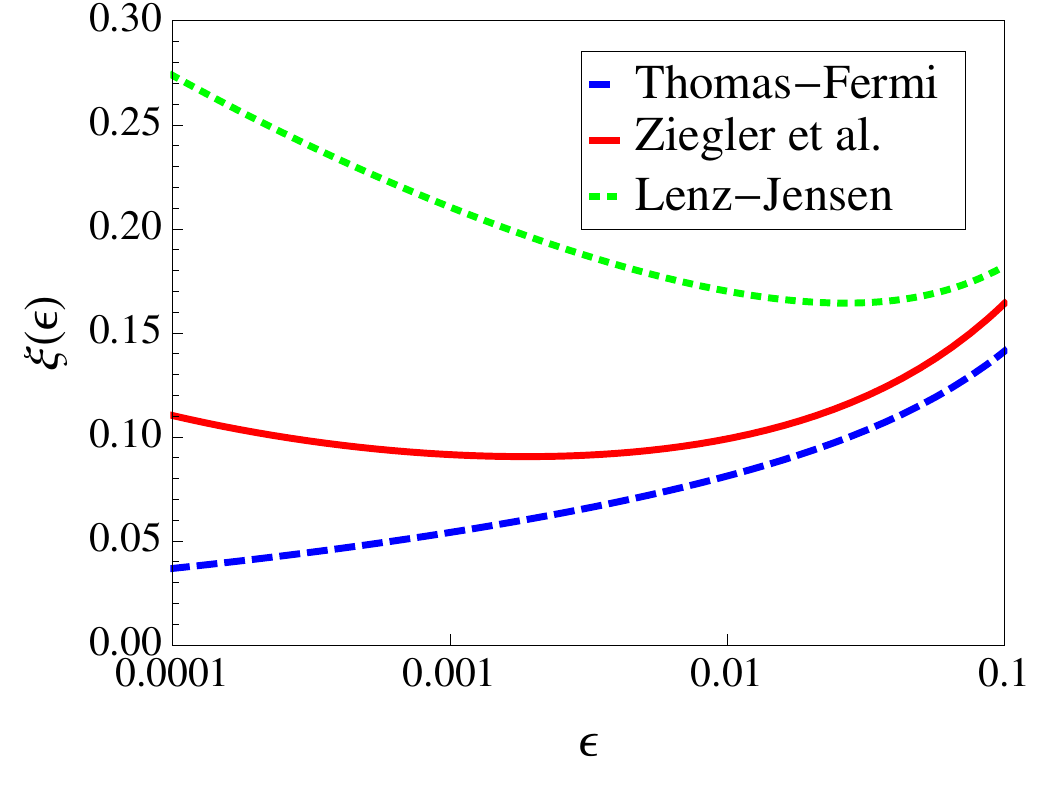}
\end{center}
\caption{The function $\xi(\epsilon)$ for different choices of the nuclear stopping power. 
The electronic stopping power was calculated using $F(v/v_0) = 1$.}
\label{fig:xi}
\end {figure}

For Thomas-Fermi screening, we obtain from \eq{f_scaling} that $\xi(\epsilon) \propto \epsilon^{0.17}$ and therefore $\kappa(\epsilon) \propto \epsilon^{1.17}$. Consequently, Lindhard's theory predicts an increasing $\xi(\epsilon)$ as the nuclear recoil energy increases from $\unit[1]{keV}$ to $\unit[100]{keV}$. This result has often been quoted as a possible explanation for the energy dependence of the scintillation yield in liquid xenon. However, we understand now that this result strongly depends on our choice for the nuclear stopping power. As argued in section \ref{sec:nuclear_stopping}, choosing a Thomas-Fermi screening function will tend to overestimate the nuclear stopping power at low energies. Consequently, we must expect to underestimate $\xi(\epsilon)$. It appears much more reasonable to choose a nuclear stopping power that agrees better with experimental data (see also \cite{Mangiarotti:2006ye}). In fact, choosing the universal stopping power from Ziegler et al.\ which is still closest to Thomas-Fermi, already changes the behavior of $\xi(\epsilon)$ considerably (see figure~\ref{fig:xi}). Now, $\xi(\epsilon)$ is no longer increasing monotonically, but develops a minimum around a few keV, remaining almost constant in most of the region we are interested in.

As we have seen, the uncertainties concerning the fraction of energy deposited in electronic excitations remain quite large. Although Lindhard's theory has been quoted frequently in the context of the effective scintillation yield, it appears difficult to obtain even a general tendency from this theory. We therefore believe that the underlying assumptions on the electronic and nuclear stopping powers are too weak to support the usual conclusion, that the energy dependence of $\Leff$ is due to a general suppression of electronic excitations at low energy. 
For this reason, a precise calculation of the relative scintillation efficiency is very difficult. To make progress, we will need to extract additional predictions from our theory that can be compared to experimental data to constrain our model. As we will show below, a reasonable agreement with experimental data is achieved for an almost constant $\xi(\epsilon)$ at low energies.

\section{Recombination}
\label{sec:recombination}

Now that we have an estimate of the total energy in electronic excitations, we need to determine how this energy is distributed between ionization and scintillation. This distribution depends not only on the number of excited and ionized atoms produced initially, but especially on the recombination rate. Recombination will occur whenever an electron and a ion produced in a nuclear recoil process approach sufficiently close. The recombination rate should be proportional to the ionization density, which in turn is roughly  proportional to the electronic stopping power $s_e(\epsilon)$. Consequently, we expect a higher ionization density, and thus a higher recombination rate, at higher recoil energies. In this section, we will follow closely the Ph.D. thesis of Dahl~\cite{Dahl}.

As discussed in section~\ref{sec:production}, after all recoiling atoms have thermalized, we are left with a certain number of excitons, called $N_\mathrm{ex}$, and a certain number of ionized atoms, $N_\mathrm{i}$. We expect that a fraction $r$ of the ionized atoms will recombine with free electrons, forming excitons that will eventually emit scintillation photons. The number of photons produced should consequently be given by\footnote{Strictly speaking, this equation describes the number of excited atoms after recombination. The actual number of photons produced will be reduced by quenching effects. However, for the moment, we neglect these effects, but will include them later on.}
\beq N_\mathrm{ph} = N_\mathrm{ex} + r \cdot N_\mathrm{i} = N_\mathrm{i} \left(r + \frac{N_\mathrm{ex}}{N_\mathrm{i}}\right) \; . \eeq
We assume that the efficiency for the production of a scintillation photon in a recombination process is close to $100\%$. Moreover, we assume that the fraction $N_\mathrm{ex} / N_\mathrm{i}$ is energy independent (see \cite{ISI:A19612681B00011} for a discussion), although it may depend on the nature of the recoiling particle.\footnote{For example, we expect $N_\mathrm{ex} / N_\mathrm{i}$ to be larger for the collision of two xenon atoms than for electron recoils, because the xenon atoms can temporarily form molecular orbitals that enhance the probability for excitations \cite{Fano:1965zz, ISI:A1976CD15400013, ISI:A1989CY20300003}.} This assumption is supported by computer simulations~\cite{Dahl} (see also ~\cite{Manalaysay} for a discussion on ways to determine $N_\mathrm{ex} / N_\mathrm{i}$ experimentally).\footnote{An analysis of experimental data \cite{Sorensen:2011bd}, which appeared after this work was first submitted, further supports this assumption.} Consequently, an energy dependence can only be introduced by the recombination fraction $r$. For later uses, we also define the number of electrons produced,
\beq N_\mathrm{q} = (1 - r) \cdot N_\mathrm{i} \;. \eeq

$N_\mathrm{ex}$ and $N_\mathrm{i}$ are presently unknown. However, they should both be proportional to $\kappa(\epsilon)$, which in turn was determined to be proportional to $\epsilon\xi(\epsilon)$. As we kept the constant of proportionality undetermined, we can do the same thing for $N_\mathrm{ex} + N_\mathrm{i}$, writing simply
\beq N_\mathrm{ex} + N_\mathrm{i} = N_\mathrm{i} \left(1 + \frac{N_\mathrm{ex}}{N_\mathrm{i}}\right) = \beta \epsilon \xi(\epsilon) \;. \eqn{Ni} \eeq
Here we have assumed that the mean energy for an excitation is approximately equal to the mean energy for an ionization, which means that $N_\mathrm{ex} / N_\mathrm{i} \approx 1$. We will see below, that this assumption is true to very good approximation. \Eq{Ni} allows to calculate $N_\mathrm{ex}$ and $N_\mathrm{i}$ from 
$\kappa(\epsilon)$ once we have determined $N_\mathrm{ex} / N_\mathrm{i}$ and $\beta$. If we also know the recombination fraction $r(\epsilon)$, we can then infer $N_\mathrm{ph}$.

\begin{figure}
\begin{center}
  \includegraphics[width=0.8\columnwidth]{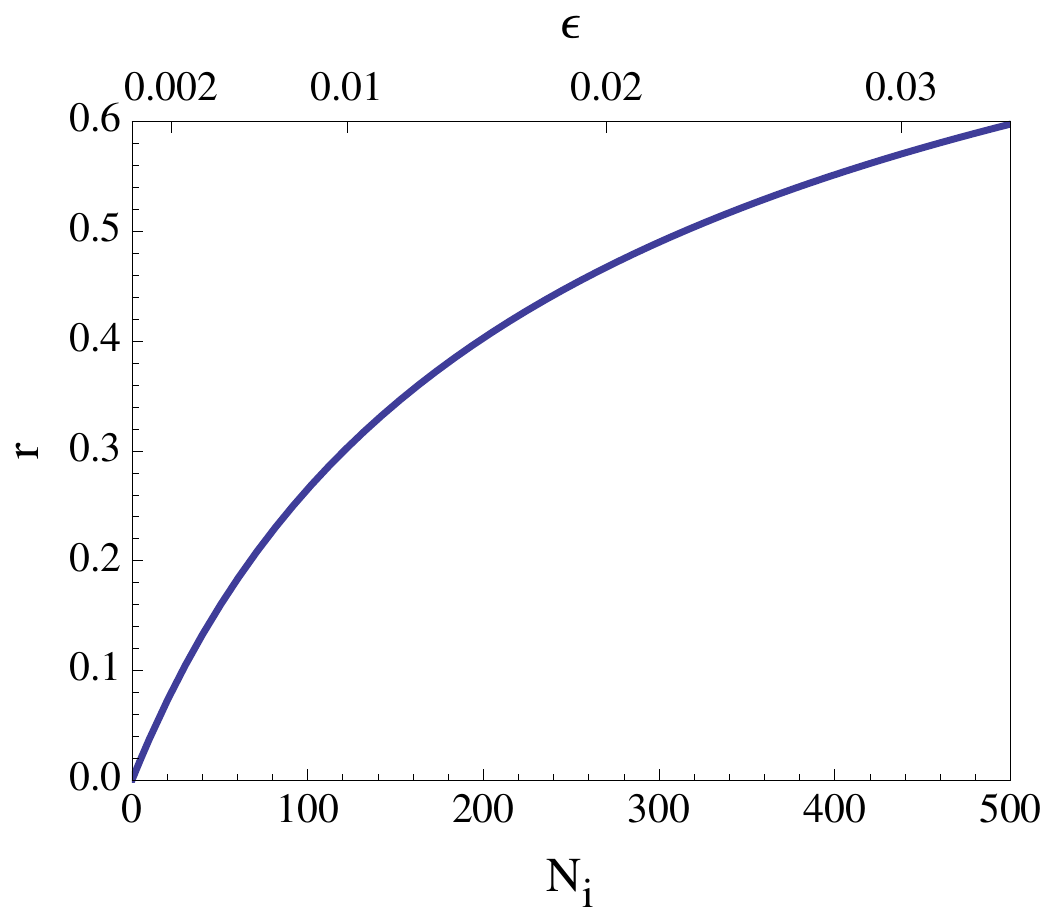}
\end{center}
\caption{Recombination fraction $r$ as a function of the number of ionized atoms. Using the results from section~\ref{sec:scintillation}, these values can be converted into reduced recoil energies, which are shown on the top of the graph.}
\label{fig:recombination}
\end{figure}

The task is therefore to determine $N_\mathrm{ex} / N_\mathrm{i}$ and $r(\epsilon)$ from an analytical model. In order to calculate the recombination rate, one needs to describe diffusion processes for electrons and ions. In fact, various theories describing recombination exist (for a review, see \cite{Amoruso:2004dy}). A model by Thomas and Imel \cite{ThomasImel} has been successfully used to describe recombination in liquid xenon \cite{Dahl,Angle:2011th, Sorensen:2011bd}. This model gives
\beq \frac{N_\mathrm{q}}{N_\mathrm{i}} = 1 - r = \frac{4}{\gamma N_\mathrm{i}} \ln \left(1 + \frac{\gamma N_\mathrm{i}}{4} \right) \;, \eqn{Nq} \eeq
where $\gamma$ is a free parameter of the theory.\footnote{In fact, $\gamma$ can be expressed in terms of the recombination coefficient, $\alpha$, the drift velocity, $v$, and the typical size of the track, $a$: $\gamma = \alpha / a^2 v$. However, as these parameters are unknown, we may just as well take $\gamma$ as the parameter of the theory.}

The two free parameters $\gamma$ and $N_\mathrm{ex} / N_\mathrm{i}$ can be determined either experimentally \cite{Sorensen:2011bd} or from Monte Carlo simulations \cite{Dahl}. Here we take the values recently proposed by the XENON10 collaboration \cite{Angle:2011th} as a conservative fit for the ionization yield in liquid xenon:
\beqs
\frac{N_\mathrm{ex}}{N_\mathrm{i}} & \approx & 1.09 \; ,\\
\gamma & \approx & 0.032 \;. \eeqs
In \cite{Sorensen:2011bd} it was argued that $N_\mathrm{ex} / N_\mathrm{i}$ has an uncertainty of about $15\%$. In addition, $\gamma$ can vary by as much as $20\%$ for different values of the electric field \cite{Dahl} in such a way that the recombination decreases with increasing field strength. As these uncertainties do not significantly affect our conclusions, we take the value from above at all electric fields in order to calculate the recombination fraction $r$ according to \eq{Nq} (figure \ref{fig:recombination}).

\section{Obtaining the scintillation yield}
\label{sec:scintillation}

\begin{figure}
\begin{center}
  \includegraphics[width=0.95\columnwidth]{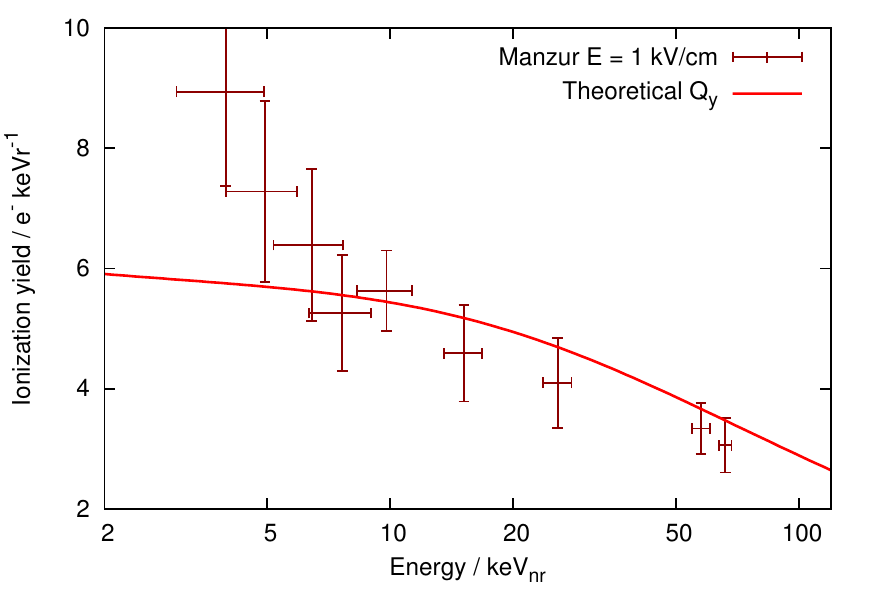}
\end{center}
\caption{Experimental data for the ionization yield from~\cite{Manzur:2009hp} for an electric field of $E = \unit[1]{kV/cm}$ and the best fit of \eq{FitNq} using $N_\mathrm{ex}/N_\mathrm{i} = 1.09$ and $\gamma = 0.032$. $\xi(\epsilon)$ was calculated using $F(v/v_0) = 1$ and the universal stopping power from Ziegler et al.}
\label{fig:ionization}
\end{figure}

Having calculated the total energy in electronic excitations and the recombination fraction, we are now able to predict both the ionization and the scintillation yield. However, one free parameter still remains in our theory: the proportionality factor $\beta$ which we introduced in \eq{Ni}. Combining this equation with \eq{Nq}, we can write
\beqs N_\mathrm{q}(\epsilon) & = & \frac{4}{\gamma} \ln \left(1 + \frac{\gamma N_\mathrm{i}(\epsilon)}{4} \right) \; ,\eqn{FitNq} \\
N_\mathrm{i}(\epsilon) & = & \frac{\beta \epsilon \xi(\epsilon)}{1 + N_\mathrm{ex} / N_\mathrm{i}} \; .
\eeqs
$N_\mathrm{q}(\epsilon)$ has been measured experimentally (most recently in~\cite{Manzur:2009hp, Horn:2011}), so we can determine $\beta$ from fitting \eq{FitNq} to the available data. Instead of $N_\mathrm{q}(\epsilon)$, one conventionally plots the ionization yield, which is defined as $Q_\mathrm{y}(E_\mathrm{nr}) = N_\mathrm{q}(E_\mathrm{nr}) / E_\mathrm{nr}$ and measured in $\unit{e^- / keVnr}$. Although there is only one free parameter, a reasonable fit can be obtained setting $\beta = (1.38 \pm 0.10) \cdot 10^5$ (see figure \ref{fig:ionization}), which indicates that our description is sufficient.\footnote{The reason why $\beta$ is so large is that $\kappa(\epsilon)$ is a reduced energy.}

In principle, a better fit could be obtained by taking $\gamma$ as a free parameter and determine it from a fit to the data. In this case, larger values of both $\beta$ and $\gamma$ are preferred, corresponding to a larger ionization yield at lower energies and a steeper decrease towards higher energies. Doing so would also lead to larger values for $\Leff$ at all energies, thus giving a rather optimistic prediction of the relative scintillation efficiency. We will therefore keep $\gamma = 0.032$ to give a conservative estimate of $\Leff$, but point out that any attempt to improve the fit in figure \ref{fig:ionization} would actually lead to an increase of $\Leff$ and not a decrease, as one might naively expect. 

We are now in the position to predict the relative scintillation efficiency. The value of $N_\mathrm{ph}$ can be obtained directly from $N_\mathrm{q}$ because $N_\mathrm{q} + N_\mathrm{ph} = N_\mathrm{ex} + N_\mathrm{i} = \beta \epsilon \xi(\epsilon)$ and this sum is known once we have determined $\beta$. In order to obtain $\Leff$, we need to divide $N_\mathrm{ph}$ by the number of photons produced by the reference electron recoil at $\unit[122]{keV}$, $n_\mathrm{ph}^\mathrm{ref}$. This value can be determined from the $W_\mathrm{ph}(\beta)$ value for xenon, which is the energy that an electron recoil must on average deposit in the detector to produce a scintillation photon.\footnote{Note that $W_\mathrm{ph}(\beta)$ is different from  $W_\mathrm{ph}$, which is the mean energy required to produce either an excited or an ionized atom, corresponding to the theoretical mean energy to produce a photon if there is full recombination and no quenching. $W_\mathrm{ph}(\beta)$ takes into account these effects and is therefore larger than $W_\mathrm{ph}$. For a careful discussion of this issue see \cite{Aprile2010RvMP}.} 

Unfortunately, $W_\mathrm{ph}(\beta)$ is known only with some uncertainty. For electrons with an energy of $\unit[1]{MeV}$, one observes $W_\mathrm{ph}(\beta) = \unit[21.6]{eV}$~\cite{Aprile2010RvMP}. However, according to \cite{Yamashita2004692}, the scintillation yield is energy dependent, with smaller values of $W_\mathrm{ph}(\beta)$ at lower energy of the electron recoil. At $\unit[122]{keV}$, the scintillation yield is measured to be about $10\%$ larger than at $\unit[1]{MeV}$, corresponding to $W_\mathrm{ph}(\beta) \approx \unit[19]{eV}$. We will use this value for our analysis and therefore $n_\mathrm{ph}^\mathrm{ref} \approx \unit[53]{ph / keV}$, but would like to emphasize that the uncertainty related to this value is quite large -- at least $10\%$. 

Before giving our result for $\Leff$ we need to consider one more process, that has been neglected so far. As mentioned in section \ref{sec:production}, the number of excitons can be reduced by biexcitonic quenching (see \cite{Hitachi:2007zz, ISI:A1992JX92900021, Hitachi:2005ti}). The idea is that in collisions of two excited atoms, only one scintillation photon is produced. Several authors have suggested a parameterization of this process in terms of Birk's saturation law \cite{Mei:2007jn, Tretyak:2009sr}. They introduce an energy dependent quenching factor $q_\mathrm{el}$ given by
\beq q_\mathrm{el} = \frac{1}{1+k \cdot s_e(\epsilon)} \; , \eqn{birk} \eeq
where $k$ is called Birk's constant and has been determined in \cite{Mei:2007jn} to be $k = \unit[2.015 \cdot 10^{-3}]{g / MeV \, cm^2} = 21.4$ in reduced units. The value suggested in \cite{Tretyak:2009sr} is smaller by about $\unit[15]{\%}$. 
Although it has been argued \cite{Hitachi:2005ti} that the parameterization in \eq{birk} is not accurate in liquid xenon, the description is sufficient to note that biexcitonic quenching is only efficient at high recoil energies, when the density of excited atoms is large. From \eq{sered}, $s_e(\epsilon) = 0.166 \sqrt{\epsilon}$, so $k \cdot s_e(\epsilon) \lesssim 1$ in the entire energy region considered and $k \cdot s_e(\epsilon) \lesssim 0.25$ below \unit[5]{keV}. When only few excited atoms are produced, biexcitonic collisions are rare and cease to reduce the photon yield. Consequently, \eq{birk} should be a sufficiently good description in the energy region we are interested in. This statement is especially true if there is a mechanism that suppresses electronic excitations at low energy, corresponding to $F(v/v_0) < 1$ for small $v$.

\begin{figure}[p]
\centering
\subfloat[The ionization yield]{\centerline{\includegraphics[width=1.05\columnwidth]{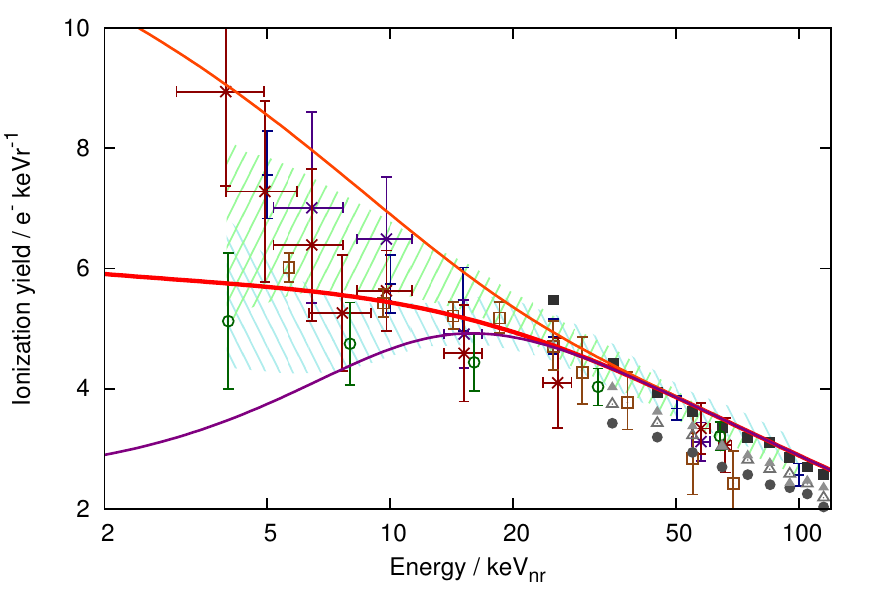}\label{fig:severalqy}}}
\qquad
\subfloat[The relative scintillation efficiency]{\centerline{\includegraphics[width=1.05\columnwidth]{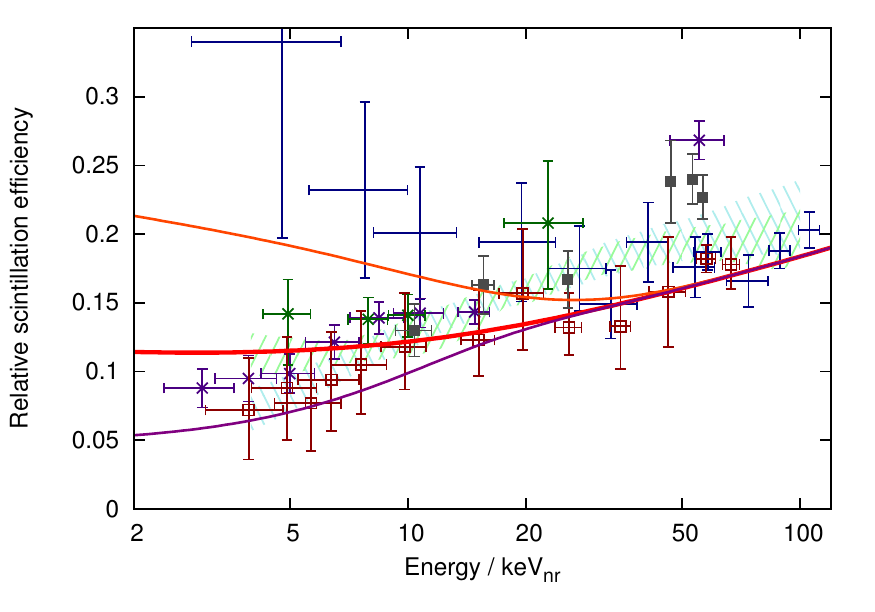}\label{fig:severalleff}}}
\caption{Predictions for the ionization yield $Q_\mathrm{y}$ and the relative scintillation efficiency $\Leff$ compared to the experimental data presented in~\cite{Horn:2011} (and further references therein). The lines with different colors correspond to different choices of the function $F(v/v_0)$: The red line, which we consider as the best description, corresponds to our proposal to modify Lindhard's theory by using the universal nuclear stopping power from Ziegler et al.\ (see also figure \ref{fig:xi}) and setting $F(v/v_0) = 1$. The orange line has been obtained by assuming an enhancement of the electronic excitations at low energies as in \eq{Fenh}. The purple line on the other hand corresponds to a suppression of the electronic stopping power by introducing a smooth cut-off for $\xi(\epsilon)$ as in \eq{Fsupp}. While the enhancement is excluded by the data for the relative scintillation efficiency, in the case of a suppression the ionization yield becomes clearly inconsistent with data, thus limiting the scintillation efficiency from below. For all plots we have used $\beta = 1.38 \cdot 10^5$.}
\label{fig:comparision_models}
\end{figure}

Now we can write down our final result for $\Leff$:
\beq \Leff = \frac{N_\mathrm{ph}(E_\mathrm{nr})}{E_\mathrm{nr} \cdot n_\mathrm{ph}^\mathrm{ref}} \cdot q_\mathrm{el}(E_\mathrm{nr}) \;. \eeq
All parameters appearing in this equation have either been measured previously or were fixed above. Consequently, we can now plot $\Leff$ and compare it with available data. For $F(v/v_0) = 1$ our model predicts a flat $\Leff$ at low recoil energies, giving roughly  $\Leff = 0.09$ at $E_\mathrm{nr} = \unit[2]{keV}$ (see the red curve in figure \ref{fig:severalleff}). Also, good agreement in both $Q_\mathrm{y}$ and $\Leff$ is found between our results and the results that Sorensen obtained from Monte Carlo simulations of the nuclear recoil band \cite{Sorensen:2010hq}.

We would like to emphasize that the Lindhard factor, meaning $\xi(\epsilon)$, is still the dominating uncertainty of our model. In fact, we are under the impression, that this uncertainty has often been underestimated previously, since not even the general trend (suppression or enhancement) of electronic excitations at low recoil energies is fully clear. However, the point we would like to emphasize is that different assumptions for the nuclear and electronic stopping powers affect the predictions for both ionization yield and relative scintillation efficiency in a correlated way.

To illustrate this point, we also show in figure \ref{fig:comparision_models} additional curves corresponding to different assumptions on $\xi(\epsilon)$ (corresponding to different choices of the correction factor $F(v/v_0)$ in \eq{suppfactor}). In detail, the cases considered are
\beqs 
F_\text{enh}(v/v_0) & = & 1 + \exp(-50\epsilon) \eqn{Fenh} \\
F_\text{supp}(v/v_0) & = & \frac{1}{2}(1 + \tanh(50\epsilon-0.25)) \eqn{Fsupp}
\eeqs
corresponding to an enhancement and a suppression of the electronic excitation at low recoil energies, respectively.\footnote{Note that
  $\epsilon$ is of course a function of $v/v_0$.} $F_\text{supp}(x)$ is similar to the result obtained by Tilinin~\cite{ISI:A1995QT40100056} in an attempt to include Coulomb effects in the calculation of the electronic stopping power. It could, in principle, account for the drop of the relative scintillation efficiency observed experimentally in~\cite{Lebedenko2009}. $F_\text{enh}(x)$ is harder to motivate from a theoretical perspective, but could result from smaller nuclear stopping power at low energies. 

Due to recent measurements of $\Leff$ \cite{Plante:2011hw}, we can clearly exclude the possibility of an enhancement of the electronic excitations at low energies as described by $F_\text{enh}(x)$. However, the remaining two predicted curves for $\Leff$ are similarly compatible with experimental data. From the theoretical point of view, it would not be easy to justify a definite preference for one of these choices either. However, choosing $F(v / v_0) < 1$ will also significantly reduce the ionization yield. If $\xi(\epsilon)$ decreases with decreasing recoil energy (as it would be the case in the presence of threshold effects or Coulomb effects), one cannot account for the increasing ionization yield that we observe experimentally. Our central observation is therefore that any model attempting to explain the scintillation yield of liquid xenon, must at the same time explain the ionization yield. Consequently, a general suppression of electronic excitations at low recoil energies is clearly incompatible with experiments. In contrast, a nearly constant value of $\xi(\epsilon)$ at low energies can accommodate (and even predict) the opposing trends seen in scintillation and ionization yield, because of the energy dependent recombination fraction.

\begin{figure}[tb]
\centering
\subfloat[The ionization yield]{\includegraphics[width=0.95\columnwidth]{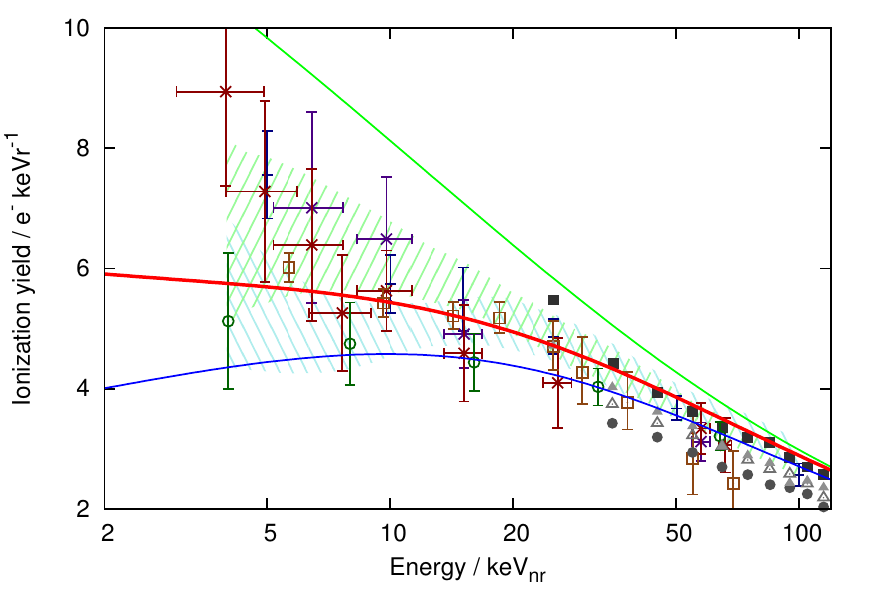}\label{fig:severalqy_stopping}}

\subfloat[The relative scintillation efficiency]{\includegraphics[width=0.95\columnwidth]{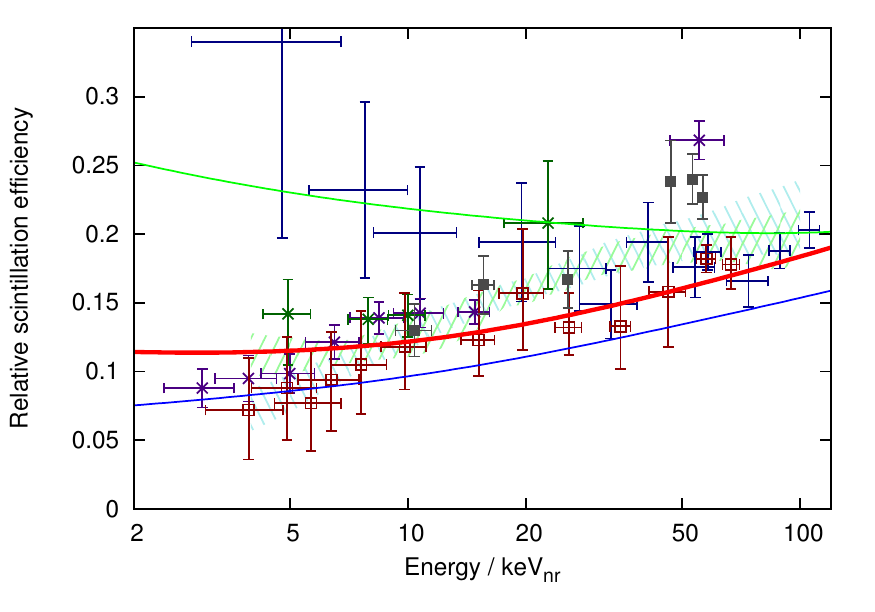}\label{fig:severalleff_stopping}}
\caption{Predictions for the ionization yield $Q_\mathrm{y}$ and the relative scintillation efficiency $\Leff$ compared to the experimental data presented in~\cite{Horn:2011} (and further references therein). The lines with different colors correspond to different choices of the nuclear stopping power as in figure~\ref{fig:xi}. For all plots we have used $\beta = 1.38 \cdot 10^5$.}
\label{fig:stopping}
\end{figure}

There are several sources of errors in the present analysis.  The main theoretical errors come out of the calculation of the nucelar and electronic stopping powers and the understanding of the ionization process.  The first, corresponding to the nuclear stopping power, is comparably small, as far as the knowledge of elastic scattering is quite good. To emphasize this point, we also show in figure~\ref{fig:stopping} the effect of taking different nuclear stopping powers to calculate the electronic excitations $\xi(\epsilon)$. Indeed, the universal stopping power from Ziegler et al.\ is not only best motivated theoretically, but also gives the best description of both ionization and scintillation.
The uncertainties in the electronic stopping powers and ionisation process, on the other hand, can change the result considerably.  Lacking a reliable calculation of xenon inelastic scattering at low energies, we cannot provide an exact estimate of the induced theoretical error~--- the wide spread of the graphs in Fig.~\ref{fig:comparision_models} can
be considered as a spread of theoretical predictions.  However, after choosing the value for $\xi(\epsilon)$ from the ionisation yield measurements (upper plot), the respective curve on the lower plot will have much smaller error.

Our results rely on the assumption that the ratio $N_\mathrm{ex} / N_\mathrm{i}$ does not vary strongly with energy. If we allow for an arbitrary energy dependence of this quantity and simultaneously vary $\xi(\epsilon)$, we could obviously fit any measurement for the ionization yield and the relative scintillation efficiency. However, from all available data, such an energy dependence is not expected. Moreover, to suppress scintillation and enhance ionization at low energies, $N_\mathrm{ex} / N_\mathrm{i}$ would have to decrease, implying that excitation becomes less likely compared to ionization. Such a behavior would most likely contradict the fact that less energy is required to excite a xenon atom than to ionize it.

\section{Conclusions}
\label{sec:conclusions}

We studied in this paper the behavior of low energy nuclear recoils in liquid xenon which is important for Dark Matter searches. An ab-initio analysis of the problem would require a quantum mechanical calculation of exclusive cross sections in Xe--Xe collisions (elastic, ionizing and leading to excited atoms) and a Monte Carlo simulation of the propagation of the nuclear recoil and further recombination. The collision analysis should be made for low impact velocities, taking into account the electronic structure of the outer shells of xenon (for example, using Density Functional Theory). Alternatively, these cross sections could be determined in experiments on scattering of individual xenon atoms.  

To do so is obviously a formidable task and we pursued therefore a simplified effective framework. Specifically we combined  Lindhard's theory for the initial production of electron excitations in atomic collisions with the well motivated assumption of energy independent partition between ionization and excitation in the underlying collisions. Including an analysis of the recombination processes we obtained \emph{both} the ionization and scintillation yields which show a correlated functional behavior when the stopping powers at low energies are varied. 

We argued that this correlated behavior allows an interesting consistency check when ionization and scintillation are measured simultaneously. We also argued that this correlation allows to use low recoil energy data for ionization to predict the low recoil energy dependence of scintillation. Using existing low recoil energy data for ionization we showed that it favors a constant behavior of $\Leff$ for low energies and that it allows to exclude the possibility of $\Leff$ dropping rapidly to zero below some threshold of $5$~keV or more. In the analysis of the first XENON100 data \cite{Aprile:2010um}, a constant behavior of $\Leff$ at low energies was assumed, which led to discussions in the literature about the reliability of the bounds for low WIMP masses. Our study shows that a constant $\Leff$ is strongly favored by the ionization data and this strengthens therefore the low WIMP mass bounds found by the XENON100 collaboration. 

\section{Acknowledgments}

F.K. thanks Rafael Lang for helpful comments and discussions.  We thank Teresa Marrodan-Undagoitia and Laura Baudis for carefully reading the draft and their useful comments. 
This work has been supported in part by the DFG Grant No. SFB-TR27 ``Neutrinos and Beyond.'' F.B. is partially sponsored by the Humboldt Foundation. F.K. is supported by the DAAD.

\bibliographystyle{elsarticle-num}


\end{document}